\documentclass[showpacs,twocolumn,prb]{revtex4}
\usepackage{graphicx}
\newcommand{\be}{\begin{equation}}
\newcommand{\ee}{\end{equation}}
\newcommand{\bea}{\begin{eqnarray}}
\newcommand{\eea}{\end{eqnarray}}
\newcommand{\bwt}{\begin{widetext}}
\newcommand{\ewt}{\end{widetext}}
\newcommand{\U}{{\cal U}}
\begin{document}
\title{Hidden Symmetries of Electronic Transport in a Disordered One-Dimensional Lattice}
\author{Wonkee Kim$^1$, L. Covaci$^1$, and F. Marsiglio$^{1-3}$}
\affiliation{ $^1$Department of Physics, University of Alberta,
Edmonton, Alberta, Canada, T6G~2J1 \\
$^2$DPMC, Universit\'e de Gen\`{e}ve, 24 Quai Ernest-Ansermet,
CH-1211 Gen\`{e}ve 4, Switzerland, \\
$^3$National Institute for Nanotechnology, National Research Council
of Canada, Edmonton, Alberta, Canada, T6G~2V4}
\begin{abstract}
Correlated, or extended, impurities play an important role in the
transport properties of dirty metals. Here, we examine, in the
framework of a tight-binding lattice, the transmission of a single
electron through an array of correlated impurities. In particular we
show that particles transmit through an impurity array in identical
fashion, regardless of the direction of transversal. The
demonstration of this fact is straightforward in the continuum
limit, but requires a detailed proof for the discrete lattice. We
also briefly demonstrate and discuss the time evolution of these
scattering states, to delineate regions (in time and space) where
the aforementioned symmetry is violated.
\end{abstract}

\date{\today}
\maketitle

\section{introduction}

In a one-dimensional lattice, correlated impurities or defects may
give rise to extended electronic states for particular energy
levels.\cite{dunlap,wu1,wu} Various disordered systems with
correlation have been investigated, such as the random
dimer,\cite{dunlap,wu} the random trimer,\cite{giri} non-symmetric
dimers,\cite{lavarda} the random trimer-dimer,\cite{farchioni} the
Thue-Morse lattice,\cite{chakrabarti} and the random polymer
chain.\cite{liu} A built-in internal structure of the impurity
configuration is important \cite{dunlap,wu} even if an internal
symmetry is not necessary.\cite{lavarda}
In the context of electronic
transport, the transmission resonance for a certain energy value
implies the disordered system behaves like an ordered lattice for
the corresponding electronic state, which is extended throughout the
entire lattice. In this respect, it is possible to understand the
increase in the conductivity for a conducting polymer\cite{wu1}.

\begin{figure}[tp]
\begin{center}
\includegraphics[height=2.6in,width=3.in]{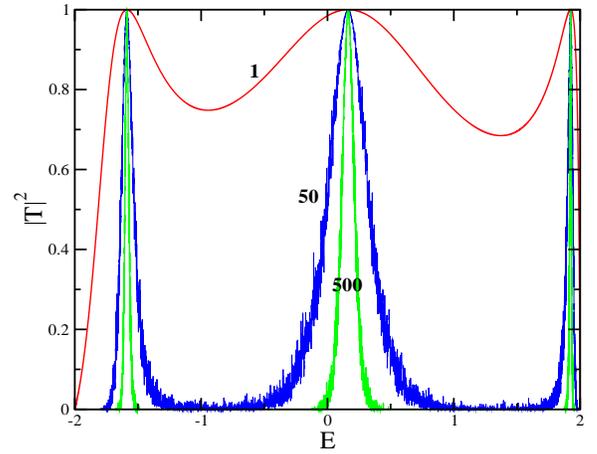}
\caption{(Color online) Transmission probability as a function of
$E$ for randomly distributed $50$ impurity dimers (blue) and $500$
dimers (green) in comparison with a single dimer (red). The inner
spacing of a pair $n_{d}=3$ and the potential $V=1/2$ while the
critical value $V_{0}=2/3$. The transmission probability is
calculated as an ensemble average over $50$ impurity configurations.
}
\end{center}
\end{figure}

One may, in fact, envision many other impurity configurations that
give rise to extended states for various energy values. For example,
let us consider a pair of impurities with a potential $V$ with
respect to the lattice potential (taken to be zero). We assume that
the spacing between the two impurities is always $n_{d}a$, where $a$
is the lattice constant and $n_{d}$ is a positive integer. This is
then a `dimer', whose length can be any value. If we adopt a
tight-binding model, so that energies are given by
$E=-2t_{0}\cos(ka)$, where $t_{0}$ is the nearest-neighbor hopping
amplitude, then one can obtain the transmission probability
$|T|^{2}$ for this state. It is \be |T|^{2} = \frac{\sin^{2}(k)}
{\sin^{2}(k)+V^{2}\left[\cos(kn_{d})+\frac{V}{2}
\frac{\sin(kn_{d})}{\sin(k)}\right]^{2}}\;.%
\label{Tr_2}%
\ee%
where $t_{0}$ and $a$ have been set to unity for simplicity.
Inspection of Eq.~(\ref{Tr_2}) reveals a critical strength of the
impurity potential $V_{c}/t_0 =\pm 2/n_{d}$ for repulsive and
attractive interactions, respectively, which determines how many
states are extended, i.e. have unit transmission. A graphical
construction readily shows that, for $V>2/n_{d}$, there are
$n_{d}-1$ extended states while for $V\le 2/n_{d}$, there are
$n_{d}$ extended states. As in the nearest-neighbor dimer case,
consideration of one `dimer' helps us to understand electronic
transport in a lattice with many `dimers', following the analysis of
Ref.\cite{dunlap}. In Fig.~1, we show $|T|^{2}$ as a function of $E$
for randomly distributed $50$ impurity dimers (blue) and $500$
dimers (green) in comparison with a single dimer (red). The inner
spacing of a pair $n_{d}=3$ and the potential $V=1/2$. Note that in
this instance $V<V_{c}=2/3$. The transmission probability is
calculated to be an ensemble average over $50$ impurity
configurations. Remarkably, the transmission remains unity at the
three energies for which a single `dimer' has unit transmission, and
the 'width' of the resonance narrows as the number of impurities
increases.

This specific example is given by way of introduction to more
general considerations of impurity potentials on a lattice. In this
paper we will show there exist hidden symmetries underlying the
physics of electronic transport in a one-dimensional lattice with an
arbitrary impurity configuration, using the transfer matrix
formalism. We will also demonstrate the time evolution of a wave
packet by direct diagonalization of the Hamiltonian since the
transfer matrix formalism does not produce the dynamics of a wave
function. While the dynamics can be insightful, we are also hopeful
that present-day technology can spatially and temporally resolve
some of the interesting dynamics that result from our (and future)
work.

Perhaps the most general correlated impurity configuration is the
following. Suppose there are $N$ impurities
A$_{1}$A$_{2}\cdots$A$_{N}$ embedded at the sites $1$ through $N$ in
a host lattice. The corresponding impurity potentials are given by
$(U_{1},\;U_{2},\;\cdots,\;U_{N})$, where $U_{i}\; (i=1,\cdots, N)$
can be positive, zero, or negative. Our main result is that the
transmission amplitude $T$ for this impurity configuration is
identical to $T$ for its reverse configuration; namely,
$T(U_{1},\;U_{2},\;\cdots,\;U_{N})=
T(U_{N},\;U_{N-1},\;\cdots,\;U_{1})$. Given this then obviously the
equality holds for the transmission $|T|^{2}$ and reflection
$|R|^{2}$ probabilities. In other words, $|T|^{2}$ and $|R|^{2}$ are
independent of the direction of the incoming wave. This symmetry is
striking because the impurity configuration is assumed arbitrary
without possessing any parity. For example, it has been shown that
the chains of ABBABAAB$\cdots$ and BAABABBA$\cdots$ contribute
identically for electronic transport\cite{chakrabarti}. This is a
simple example of the general symmetry we describe. Note that an
exchange of any two impurity potentials does not show this symmetry;
for example, $T(U_{1},\;U_{2},\;U_{3},\;U_{4})\ne
T(U_{2},\;U_{1},\;U_{3},\;U_{4})$ \cite{wkb}.

We should emphasize now that this symmetry does {\em not} hold in
the impurity region (as perhaps one would expect), but this can only
be determined by solving the time-dependent Schr{\"o}dinger equation
(see below). We also explain another symmetry associated with a
particular momentum $k=\pi/(2a)$, where $a$ is the lattice constant:
$|T(U_{1},\;U_{2},\;\cdots,\;U_{N})|^{2}=
|T(-U_{1},\;-U_{2},\;\cdots,\;-U_{N})|^{2}$. This equality indicates
a potential barrier or well produces identical transmission and
reflection probabilities for the electronic state with $k=\pi/(2a)$.
This is unrelated to the other remarkable feature of wave packets on
a one-dimensional lattice with nearest-neighbor hopping and
$k=\pi/(2a)$: they do not diffuse as a function of time, because
they lack group velocity dispersion. Indeed, one can show that no
matter what the range of the hopping is, there will be a wave vector
for which the wave packet retains its initial width.

\section{formalism}

We start with a tight-binding Hamiltonian for a one-dimensional
lattice with $N$ impurities:
\be
H=-t_{0}\sum_{i}\left[c^{+}_{i}c_{i+1}+c^{+}_{i+1}c_{i}\right]+
\sum_{i\in {\cal I}}U_{i}c^{+}_{i}c_{i}
\ee
where the
hopping amplitude $t_{0}$ will be set to be unity, $c^{+}_{i}$
creates an electron at a site $i$, and ${\cal I}$ represents an
impurity configuration. The Schr{\"o}dinger equation is
$H|\psi\rangle = E|\psi\rangle$ with $|\psi\rangle =
\sum_{j}\psi_{j}c^{+}_{j}|0\rangle$; this becomes%
\be%
-\left[\psi_{j+1}+\psi_{j-1}\right]+U_{j}\psi_{j}=E\psi{j}%
\label{S_eq}%
\ee%
where $\psi_{j}$ is an amplitude to find an electron at site $j$. In
order to use the transfer matrix formalism we write Eq.~(\ref{S_eq})
in a matrix form as follows: \be \left(\begin{array}{c}
\psi_{j+1}\\
\psi_{j}
\end{array}\right)
=
\left(\begin{array}{cc}
U_{i}-E & -1\\
1 & 0
\end{array}\right)
\left(\begin{array}{c}
\psi_{j}\\
\psi_{j-1}
\end{array}\right)
\equiv M_{j}
\left(\begin{array}{c}
\psi_{j}\\
\psi_{j-1}
\end{array}\right)\;.
\ee Note that $M_{i}$ is a unimodular matrix, i.e. $\det(M_{i})=1$.
The wave functions $\psi_{L}$ (for $i<1$) and $\psi_{R}$ (for $i>N$)
are $\psi_{L}=e^{ikx_{i}}+R\;e^{-ikx_{i}}$ and
$\psi_{R}=T\;e^{ikx_{i}}$, where $x_{i}=a\cdot i$. Using the
transfer matrix formalism, one can express the coefficients $R$ and
$T$ in terms of $k$, $U_{i}$, and $E$ as follows: \be
\left(\begin{array}{c}
T\\
iT
\end{array}\right)
=P
\left(\begin{array}{c}
1+R\\
i(1-R)
\end{array}\right)\;,
\label{tr_eq}
\ee
where $P=S^{-1}MS$ with $S=\left(\begin{array}{cc}
\cos(k) & \sin(k)\\
1 & 0
\end{array}\right)$, and $M=M_{N}M_{N-1}\cdots M_{1}$.
It is misleading to express $M=\Pi^{N}_{i=1}M_{i}$ because $M_{i}$
and $M_{j}$ $(i\ne j)$ do not commute with each other. Solving
Eq.~(\ref{tr_eq}), one can obtain\cite{burrow}
\bea
T&=&\frac{2i}{i\left(P_{11}+P_{22}\right)+P_{12}-P_{21}}
\\
R&=&\frac{P_{12}+P_{21}-i\left(P_{11}-P_{22}\right)}
{i\left(P_{11}+P_{22}\right)+P_{12}-P_{21}}\;.
\eea
Consequently, the transmission probability
$|T|^{2}=4\big/\left[\mbox {tr}(P{\tilde P})+2\right]$ and the
reflection probability $|R|^{2}=\left[\mbox {tr}(P{\tilde
P})-2\right]\big/ \left[\mbox {tr}(P{\tilde P})+2\right]$, where
tr$(P)$ means the trace of $P$, and ${\tilde P}$ is the transpose of
$P$. Note that Eq.~(\ref{Tr_2}) is readily obtained with this
formalism by using $M'=M_{V}M^{n_{d}-1}_{0}M_{V}$, where
$M_{V}=M_{1}(U_{1}=V)$ and $M_{0}=M_{1}(U_{1}=0)$, and $n_d$ is the
inner spacing of the impurity pair.

One of the symmetries we introduced earlier is
$T(U_{1},\;U_{2},\cdots, \;U_{N})= T(U_{N},\;U_{N-1},\cdots,
\;U_{1})$. In order to show this equality, let us introduce
$Q=S^{-1}M'S$ where $M'=M_{1}M_{2}\cdots M_{N}$. Notice the order of
the matrix multiplication because $M_{i}$ and $M_{j}$ $(i\ne j)$ do
not commute with each other. Since the desired equality implies that
$T[P]=T[Q]$, we need to show $P_{11}+P_{22}=Q_{11}+Q_{22}$ and
$P_{12}-P_{21}=Q_{12}-Q_{21}$. On the other hand, from the
definitions of $P=S^{-1}MS$ and $Q=S^{-1}M'S$, this equality in turn
implies that $M_{11}=M'_{11}$, $M_{22}=M'_{22}$, and
$M_{12}-M_{21}=M'_{12}-M'_{21}$. Note, however, that in general
$M\ne M'$, as we will show later.

For simplicity, we define $\U_{i}=U_{i}-E$. We also introduce
$2\times2$ matrices $\alpha$ and $\beta$ such as
$\alpha=\left(\begin{array}{cc}
1 & 0\\
0 & 0
\end{array}\right)$ and
$\beta=\left(\begin{array}{cc}
0 & -1\\
1 & 0
\end{array}\right)$ to have $M_{i}=\U_{i}\alpha+\beta$.
Note that $\alpha^{2}=\alpha$, $\beta^{2}=-1$, $\alpha\beta+\beta\alpha=\beta$,
and $\alpha\beta\alpha=0$. Since $M=M_{N}M_{N-1}\cdots M_{1}$,
we obtain
\bwt
\bea
M&=&(\U_{N}\alpha+\beta)(\U_{N-1}\alpha+\beta)\cdots(\U_{1}\alpha+\beta)
\nonumber\\
&=&\beta^{N}+\sum^{N}_{n}\sum^{N}_{\{j_{n}\}}
(\U_{j_{1}}\U_{j_{2}}\cdots\U_{j_{n}})\beta^{N-j_{n}}
\alpha\beta^{j_{n}-j_{n-1}-1}\alpha\cdots\alpha
\beta^{j_{2}-j_{1}-1}\alpha\beta^{j_{1}-1}\;,
\eea
where $\{j_{n}\}$ means $j_{1},j_{2},\cdots,j_{n}=1,2,,\cdots,N$
with $j_{1}<j_{2}<\cdots<j_{n}$. Since $\alpha\beta\alpha=0$ and
$\beta^{2}=-1$, non-vanishing terms should have
$j_{l}-j_{l-1}-1=\mbox{even}$, where $l=2,3,\cdots n$ for a given $n$.
Therefore, we obtain
\be
M=\beta^{N}+\sum^{N}_{n}\sum^{N}_{\{\{j_{n}\}\}}
(\U_{j_{1}}\U_{j_{2}}\cdots\U_{j_{n}})(-1)^{\frac{1}{2}(j_{n}-j_{1}-n+1)}
\beta^{N-j_{n}}\alpha\beta^{j_{1}-1}\;,
\ee
\ewt
where $\{\{j_{n}\}\}$ means $\{j_{n}\}$ with $j_{l}-j_{l-1}-1=\mbox{even}$.
Since
$\alpha=\left(\begin{array}{cc}
1 & 0\\
0 & 0
\end{array}\right)$,
$\alpha\beta=\left(\begin{array}{cc}
0 & -1\\
0 & 0
\end{array}\right)$,
$\beta\alpha=\left(\begin{array}{cc}
0 & 0\\
1 & 0
\end{array}\right)$,
and
$\beta\alpha\beta=\left(\begin{array}{cc}
0 & 0\\
0 & -1
\end{array}\right)$, we use these as the basis matrices to expand
$M$:
\be
M=c_{1}\alpha+c_{2}\alpha\beta+c_{3}\beta\alpha+c_{4}\beta\alpha\beta
=\left(\begin{array}{cc}
c_{1} & -c_{2}\\
c_{3} & -c_{4}
\end{array}\right)\;.
\label{eq_M}
\ee
As we mentioned, the equality of the transmission amplitude,
$T[P]=T[Q]$, is associated with
some particular relations among the components of the two
matrices $M=M_{1}M_{2}\cdots M_{N}$
and $M'=M_{N}M_{N-1}\cdots M_{1}$.
Following the same way as for $M$, we expand $M'$ in terms of
$\U_{i}\; (i=1,2,\cdots, N)$, $\alpha$, and $\beta$
to obtain
\bwt
\bea
M'&=&(\U_{1}\alpha+\beta)(\U_{2}\alpha+\beta)\cdots(\U_{N}\alpha+\beta)
\nonumber\\
&=&\beta^{N}+\sum^{N}_{n}\sum^{N}_{\{j_{n}\}}
(\U_{j_{1}}\U_{j_{2}}\cdots\U_{j_{n}})\beta^{j_{1}-1}
\alpha\beta^{j_{2}-j_{1}-1}\alpha\cdots\alpha
\beta^{j_{n}-j_{n-1}-1}\alpha\beta^{N-j_{n}}\;.%
\eea%
It is also true
for $M'$ that non-vanishing terms have $j_{l}-j_{l-1}-1=\mbox{even}$
because $\alpha\beta\alpha=0$. Therefore,
\be%
M'= \beta^{N}+\sum^{N}_{n}\sum^{N}_{\{\{j_{n}\}\}}
(\U_{j_{1}}\U_{j_{2}}\cdots\U_{j_{n}})(-1)^{\frac{1}{2}(j_{n}-j_{1}-n+1)}
\beta^{j_{1}-1}\alpha\beta^{N-j_{n}}\;.%
\ee%
\ewt%
Expanding $M'$, again, in terms of the basis matrices, we know \be
M'=
c'_{1}\alpha+c'_{2}\alpha\beta+c'_{3}\beta\alpha+c'_{4}\beta\alpha\beta
=\left(\begin{array}{cc}
c'_{1} & -c'_{2}\\
c'_{3} & -c'_{4}
\end{array}\right)\;.
\label{eq_M'} \ee Comparing Eqs.~(\ref{eq_M}) and (\ref{eq_M'}), the
equalities $M_{11}=M'_{11}$, $M_{22}=M'_{22}$, and $M_{12}-M_{21}
=M'_{12}-M'_{21}$ are equivalent to the equalities $c_{1}=c'_{1}$,
$c_{4}=c'_{4}$, and $c_{2}+c_{3}=c'_{2}+c'_{3}$. Since $\beta^{N}$
does not depend on $\U_{i}$ and $\beta^{N}=(-1)^{N/2}$ or
$(-1)^{(N-1)/2}\beta$ for even (or odd) $N$, we can ignore
$\beta^{N}$ for the purpose of showing these equalities, or,
alternatively, we can redefine $M$ as $M-\beta^{N}$ and $M'$ and
$M'-\beta^{N}$. To obtain i) $c_{1}$ of $M$, we should have
$j_{1}-1=$ even and $N-j_{n}=$ even. Similarly, ii) for $c_{2}$,
$j_{1}-1=$ odd and $N-j_{n}=$ even, iii) for $c_{3}$, $j_{1}-1=$
even and $N-j_{n}=$ odd, and iv) for $c_{4}$, $j_{1}-1=$ odd and
$N-j_{n}=$ odd. In terms of $\U_{i}$, we obtain \bea c_{1}&=&
\sum^{N}_{n}\sum^{N}_{\{\{j_{n}\}\}_{1}}
(\U_{j_{1}}\U_{j_{2}}\cdots\U_{j_{n}})(-1)^{\frac{1}{2}(N-n)}
\nonumber\\
c_{2}&=&
\sum^{N}_{n}\sum^{N}_{\{\{j_{n}\}\}_{2}}
(\U_{j_{1}}\U_{j_{2}}\cdots\U_{j_{n}})(-1)^{\frac{1}{2}(N-n-1)}
\nonumber\\
c_{3}&=&
\sum^{N}_{n}\sum^{N}_{\{\{j_{n}\}\}_{3}}
(\U_{j_{1}}\U_{j_{2}}\cdots\U_{j_{n}})(-1)^{\frac{1}{2}(N-n-1)}
\nonumber\\
c_{4}&=&
\sum^{N}_{n}\sum^{N}_{\{\{j_{n}\}\}_{4}}
(\U_{j_{1}}\U_{j_{2}}\cdots\U_{j_{n}})(-1)^{\frac{1}{2}(N-n-2)}
\nonumber
\eea
where $\{\{j_{n}\}\}_{1}$ means $\{\{j_{n}\}\}$ with
$(j_{1}-1,N-j_{n})=$ (even, even). Similarly, $\{\{j_{n}\}\}_{2}=$
$\{\{j_{n}\}\}$ with (odd, even),
$\{\{j_{n}\}\}_{3}=$ $\{\{j_{n}\}\}$ with (even, odd),
and $\{\{j_{n}\}\}_{4}=$ $\{\{j_{n}\}\}$ with (odd, odd).

For $c'_{1}$, $c'_{2}$, $c'_{3}$, and $c'_{4}$ of Eq.~(\ref{eq_M'}),
we should have $(j_{1}-1,N-j_{n})=$ (even, even), (even, odd),
(odd, even), and (odd, odd), respectively.
Note that we need (even, odd) for $c'_{2}$ while (odd, even)
for $c_{2}$. On the other hand, we need (odd, even) for $c'_{3}$
while (even, odd) for $c_{3}$.
Therefore, we obtain
\bea
c'_{1}&=&
\sum^{N}_{n}\sum^{N}_{\{\{j_{n}\}\}_{1}}
(\U_{j_{1}}\U_{j_{2}}\cdots\U_{j_{n}})(-1)^{\frac{1}{2}(N-n)}
\nonumber\\
c'_{2}&=&
\sum^{N}_{n}\sum^{N}_{\{\{j_{n}\}\}_{3}}
(\U_{j_{1}}\U_{j_{2}}\cdots\U_{j_{n}})(-1)^{\frac{1}{2}(N-n-1)}
\nonumber\\
c'_{3}&=&
\sum^{N}_{n}\sum^{N}_{\{\{j_{n}\}\}_{2}}
(\U_{j_{1}}\U_{j_{2}}\cdots\U_{j_{n}})(-1)^{\frac{1}{2}(N-n-1)}
\nonumber\\
c'_{4}&=&
\sum^{N}_{n}\sum^{N}_{\{\{j_{n}\}\}_{4}}
(\U_{j_{1}}\U_{j_{2}}\cdots\U_{j_{n}})(-1)^{\frac{1}{2}(N-n-2)}
\nonumber
\eea
Consequently, we have $c_{1}=c'_{1}$, $c_{4}=c'_{4}$,
and $c_{2}+c_{3}=c'_{2}+c'_{3}$; in other words,
$T[P]=T[Q]$, or equivalently,
$T(U_{1},\;U_{2},\;\cdots,\;U_{N})=T(U_{N},\;U_{N-1},\;\cdots,\;U_{1})$
In fact, $M_{12}=-M'_{21}$ and $M_{21}=-M'_{12}$.
Note, however, that in general $M\ne M'$.

One may anticipate a similar symmetry in the continuum
limit.\cite{arias} Consider a time-dependent Sch{\"o}dinger
equation: $i\partial_{t}\psi(t,x)=H(x)\psi(t,x)$, where the
Hamiltonian $H(x)$ includes an impurity potential $V(x)$ for $|x|\le
l$. The wave function for $|x|>l$ can be represented as
$\psi(|x|>l)=\left[e^{ikx}+Re^{-ikx}\right]\Theta(-x-l)
+Te^{ikx}\Theta(x-l)$, where $\Theta(x)$ is the step function. Let
us introduce another time-dependent Sch{\"o}dinger equation:
$i\partial_{t}\psi'(t,x)=H(-x)\psi'(t,x)$, where $\psi(|x|>l)=
\left[e^{ikx}+R'\;e^{-ikx}\right]\Theta(-x-l)
+T'\;e^{ikx}\Theta(x-l)$. Invoking both space inversion and time
reversal, one can see that $\psi(t,-x)$ and $\psi^{*}(-t,-x)$
satisfy the same equation as $\psi'(t,x)$ does. Equating
$a_{1}\psi(t,-x)+a_{2}\psi^{*}(-t,-x)=\psi'(t,x)$ for $|x|>l$ leads
to $T'=T$ and $R'T^{*}+R^{*}T=0$. The first relation corresponds to
$T[P] = T[Q]$ in a lattice. The second relation, however, does not
hold in a lattice. Instead, we found in a lattice \be
\frac{T^{*}[P]\left(R[Q]-R[P]\right)}
{T[P]\left(R^{*}[Q]-R^{*}[P]\right)}= e^{2ik}\;. \ee Nevertheless,
we can still define $R[Q] = e^{i\delta}R[P]$. The phase $\delta$ is
determined by $e^{i\delta}=-e^{2ik}\;R^{*}T/(RT^{*})$ in a lattice
while $e^{i\delta}=-R^{*}T/(RT^{*})$ in the continuum limit.

We also found another symmetry associated with electronic transport
in a one-dimensional lattice; in this case there is no applicability
in the continuum limit. Suppose $k=\pi/2$; then
$S=\left(\begin{array}{cc}
0 & 1\\
1 & 0
\end{array}\right)$, and $S={\tilde S}=S^{-1}$.
Note that for a symmetric $(2\times2)$ matrix $X$ such that $X={\tilde X}$,
$M_{i}$ satisfies $M_{i}X{\tilde M}_{i}={\tilde M}_{i}XM_{i}$ with
$\U_{i}\rightarrow -\U_{i}$. Now let us consider $\mbox{tr}(P{\tilde P})$:
\bea
\mbox{tr}\left(P{\tilde P}\right)&=&
\mbox{tr}\left(M_{N}\cdots M_{1}{\tilde M}_{1}
\cdots{\tilde M}_{N}\right)
\nonumber\\
&=&\mbox{tr}\left({\tilde M}_{N}\cdots{\tilde M}_{1}
M_{1}\cdots M_{N}\right)_{\U\rightarrow -\U}
\nonumber\\
&=&\mbox{tr}\left({\tilde Q}Q\right)_{-\U}= \mbox{tr}\left(Q{\tilde
Q}\right)_{-\U} \eea This symmetry indicates that for $k=\pi/2$,
$|T|^{2}$ and $|R|^{2}$ depend only on $|\U_{i}|$. If $E=-2\cos(k)$,
then $\U_{i}=U_{i}$. In this instance, $|T(U_{1},\cdots,U_{N})|^{2}=
|T(-U_{N},\cdots,-U_{1})|^{2}$. Since $|T(U_{1},\cdots,U_{N})|^{2}=
|T(U_{N},\cdots,U_{1})|^{2}$ from a general derivation,
$|T(U_{1},\cdots,U_{N})|^{2}= |T(-U_{1},\cdots,-U_{N})|^{2}$.
Consequently, a potential barrier and a potential well give
identical transmission and reflection probabilities for $k=\pi/2$
and $E=-2\cos(k)$.

\section{time evolution of a wave packet}

So far we have been using the transfer matrix formalism. Since,
however, the transfer matrix formalism is based on the
time-independent Schr{\"o}dinger equation, we cannot see the time
evolution of a wave function, which may indicate the significance of
the symmetries we have shown. Let us consider a wave packet with
average position $x_{0}$ and average momentum $k_{0}$ at time $t=0$:
$|\Psi(0)\rangle = \sum_{i}\varphi(x_{i},0)c^{+}_{i}|0\rangle$,
where \be \varphi(x_{i},0)=\frac{1}{(2\pi
\alpha^{2})^{1/4}}e^{ik_{0}(x_{i}-x_{0})}
e^{-\frac{1}{4}(x_{i}-x_{0})^{2}/\alpha^{2}}\;. \ee We introduced
the initial uncertainty $\alpha$ associated with position. We wish
to propagate (in real time) this wave packet towards the potentials.
To do this we first diagonalize the Hamiltonian, which is an
$(L\times L)$ matrix, where $L$ is the total number of lattice sites
under consideration. We thus obtain the eigenstates $|n\rangle$ and
the corresponding eigenvalues $\epsilon_{n}$ such that $H|n\rangle =
\epsilon_{n}|n\rangle$. An eigenstate is a column vector with $L$
components which describes the probability of finding an electron at
a particular site in the lattice. With eigenstates in hand, the time
evolution of the wave packet is given by
\be%
|\Psi(t)\rangle =
\sum^{L}_{n=1}|n\rangle\langle n|\Psi(0)\rangle
e^{-i\epsilon_{n}t}\;.%
\ee%
As time goes on, the wave packet initially at $x_{0}$ moves to the
potential region and scatters off the potential. A part of the wave
packet is reflected and the other part is transmitted.
Mathematically we can define the reflection and the transmission as
$|R|^{2}=\sum_{left}|\varphi(x_{i},t)|^{2}$ and
$|T|^{2}=\sum_{right}|\varphi(x_{i},t)|^{2}$, respectively as
$t\rightarrow\infty$, where $\sum_{left(right)}$ means the summation
includes only the left(right) side of the impurity region.

\begin{figure}[tp]
\begin{center}
\includegraphics[height=2.55in,width=3.in]{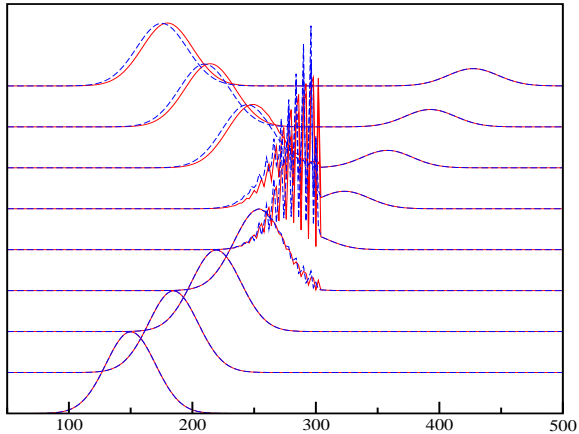}
\caption{Time evolution of a wave packet in two impurity configurations
[I] (red solid) and [II] (blue dashed curve).
Each configuration has 5 impurities: for [I] the potential is
$(-0.5,\;0,\;0.5,\;1.5,\;1)$ while for [II] its reverse holds.
}
\end{center}
\end{figure}

Fig.~2 shows the time evolution of a wave packet impinging on 5
impurities located at sites $300$ to $304$. The parameters of the
wave packet are given as follows: $x_{0}=150$, $\alpha=20$, and
$k_{0}=\pi/3$. Note we choose $\alpha=20$ so that the size of the
wave packet is much greater than the size of the impurity region.
This means that we can consider the wave packet as a plane wave in
this case. In fact, the numerical results of the transmission and
reflection probabilities are in close agreement with those obtained
through the transfer matrix analysis. The impurity configuration is
determine by 5 impurity potentials:
$(U_{1},\;U_{2},\;U_{3},\;U_{4},\;U_{5})$. For [I] (red solid), we
have $(-0.5,\;0,\;0.5,\;1.5,\;1)$ while for [II] (blue dashed
curve), we have the reverse order. As clearly shown in Fig.~2, the
time evolution of the wave packet is different depending on the
impurity configurations. In particular, the scattering completely
differentiates [I] from [II], as indicated by the very differently
behaved red and blue curves in the scattering region. Nonetheless
the probability that emerges (either in transmission or reflection)
is identical for both, in agreement with the symmetry we just
proved. Note that the transmitted wave packets are identical in all
other respects as well whereas the reflected wave packets have
identical shape, but are phase-shifted with respect to one another.
In fact, one can show that the phase shift originates from the phase
$\delta$ in $R[Q]=e^{i\delta}\; R[P]$. The shift measured by the
difference between the two reflected wave packets for [I] and [II]
at their half width is determined by $\partial\delta(k)/\partial
k\big|_{k_{0}}$.

\section{conclusions}

We have determined the transmission and reflection characteristics
for various impurity configurations on a one dimensional lattice. In
particular we proved that the transmission of a particle through an
array of impurities is independent of the direction of travel. This
theorem may help understand, among other things, weak localization,
where time-reversed paths play an important role. Some important
differences arise because of the lattice: first, it becomes clear
that the group velocity is most important (in the continuum limit
the energy is usually emphasized). Secondly, other symmetries exist
on a lattice for particular wavevectors. Naturally these symmetries
do not hold in the actual scattering region. While we have no
specific proposals at present, we hope this work motivates
experimentalists to look for these violations in the vicinity of
particular impurity configurations. Further work is in progress with
trimer impurity configurations, more general band structures, and
higher dimensionality.

\vskip 0.5cm This work was supported in part by the Natural Sciences
and Engineering Research Council of Canada (NSERC), by ICORE
(Alberta), and by the Canadian Institute for Advanced Research
(CIAR). FM is appreciative of the hospitality of the Department of
Condensed Matter Physics at the University of Geneva.

\bibliographystyle{prb}

\begin{thebibliography}{1}
\bibitem{wkb} The WKB approximation would incorrectly show this
symmetry, for example, so, while the WKB approximation clearly
satisfies the symmetry which we are about to prove, it is unreliable
as a diagnostic tool.
\end{thebibliography}

\end{document}